\documentclass[conference]{IEEEtran}

\usepackage{cite}
\usepackage{amsmath,amssymb,amsfonts}
\usepackage{graphicx}
\usepackage{booktabs}
\usepackage{hyperref}

\newcommand{\ie}{\textit{i.e.}}

\begin{document}

 \title{Towards Symmetry-Aware Efficient Simulation of Quantum Systems and Beyond}

\author{
\IEEEauthorblockN{
Min Chen\IEEEauthorrefmark{3},
Minzhao Liu\IEEEauthorrefmark{1}\IEEEauthorrefmark{2},
Changhun Oh\IEEEauthorrefmark{4}\IEEEauthorrefmark{5},
Liang Jiang\IEEEauthorrefmark{5},
Yuri Alexeev\IEEEauthorrefmark{2}$^\text{1}$, and
Junyu Liu\IEEEauthorrefmark{3}\IEEEauthorrefmark{5}}
\IEEEauthorblockA{\IEEEauthorrefmark{3}\textit{Department of Computer Science, University of Pittsburgh, Pittsburgh, PA 15260, USA}}
\IEEEauthorblockA{\IEEEauthorrefmark{1}\textit{Department of Physics, The University of Chicago, Chicago, IL 60637, USA}}
\IEEEauthorblockA{\IEEEauthorrefmark{2}\textit{Computational Science Division, Argonne National Laboratory, Lemont, IL 60439, USA}}
\IEEEauthorblockA{\IEEEauthorrefmark{4}\textit{Department of Physics, Korea Advanced Institute of Science and Technology, Daejeon 34141, Korea}}
\IEEEauthorblockA{\IEEEauthorrefmark{5} \textit{Pritzker School of Molecular Engineering, The University of Chicago, Chicago, IL 60637, USA}}

\IEEEauthorblockA{Emails: junyuliu@pitt.edu}
}

\maketitle

\begin{abstract}
The efficient simulation of complex quantum systems remains a central challenge due to the exponential growth of Hilbert space with system size. Tensor network methods have long been established as powerful approximation schemes, and their efficiency can be further enhanced by incorporating physics-informed priors. A prominent example is symmetry: recent progress on $U(1)$-symmetric tensor networks, accelerated on GPUs and scaled to supercomputers, shows how conserved charges induce block-sparse structures that reduce computational cost and enable larger simulations. The same principle extends to general symmetries, inspiring equivariant neural networks in machine learning and guiding symmetry-preserving ans\"atze in variational quantum algorithms. Beyond symmetry, physics-informed design also includes strategies such as hybrid tensor networks and parallel sequential circuits, which pursue efficiency from complementary principles. This Perspective argues that physics-informed tensor networks, grounded in both symmetry and beyond-symmetry insights, provide unifying strategies for scalable approaches in quantum simulation, computation, and machine learning.
\end{abstract}


\section{Introduction}

Quantum computing offers a fundamentally new paradigm for simulating quantum systems and tackling information processing tasks beyond the reach of classical methods. While quantum devices hold the long-term promise of efficiently tackling classically intractable many-body problems~\cite{peruzzo2014variational}, current hardware in the \textit{Noisy Intermediate-Scale Quantum (NISQ)} era~\cite{preskill2018quantum} remains limited in system size, coherence, and overall noise levels. As a result, classical simulation of quantum systems continues to play a crucial role both for benchmarking quantum processors and for exploring regimes beyond present quantum hardware. Efficiently simulating complex quantum systems, such as the quantum many-body systems, is a long-standing challenge because the exponential growth of Hilbert space with system size makes their simulation challenging on classical computers. Tensor network methods have emerged as one of the most powerful approximation schemes to tackle this challenge~\cite{li2024unifying,liu2023supercomputing,zheng2023speeding,white1992density,vidal2003entanglement,verstraete2004renormalization,liu2023simulating}. By exploiting the entanglement structure of low-energy states, tensor networks such as matrix product states (MPS)~\cite{fannes1992finitely,cirac2021matrix,schollwock2011density} and projected entangled pair states (PEPSs)~\cite{verstraete2004renormalization,verstraete2008matrix} provide efficient representations of quantum states that would otherwise be intractable. 
While these methods already capture a wide class of many-body phenomena, their efficiency can be further improved through physics-informed strategies. One is through the explicit incorporation of physical symmetries into the tensor network formalism. 
A prominent example is the use of global \(U(1)\) symmetry~\cite{singh2011tensor,singh2012tensor}, which naturally arises in models such as the Bose--Hubbard Hamiltonian~\cite{aizenman2004bose}, the XXZ spin chain~\cite{alcaraz1989xxz}, and boson sampling circuits~\cite{aaronson2011computational}. Building on this representative case, we extend our focus to the broader theme of symmetry-enhanced tensor networks, highlighting their potential impact on scalable quantum simulation. As a concrete example, we first discuss \(U(1)\)-symmetric implementations, which provide a representative case. Then the scope is naturally extended to more general Abelian and non-Abelian symmetries. We further connect these developments to parallel advances in machine learning where symmetry and equivariance are emerging as unifying design principles, as well as to variational quantum algorithms that similarly exploit structural priors. At the same time, we also acknowledge complementary directions that pursue efficiency beyond symmetry, such as hybrid tensor networks and circuit layouts tailored for noise robustness. Together, these efforts illustrate a broader landscape of principles including symmetry and symmetry-beyond that can guide the design of scalable quantum simulation, quantum computation and machine learning.

\section{Symmetry in tensor networks for quantum simulation}

Incorporating symmetries into tensor network algorithms provides a powerful way to further reduce computational complexity beyond entanglement-based compression alone. 
Global symmetries restrict the accessible subspaces of the Hilbert space, which can substantially reduce computational cost. As an illustrative example, we focus on the explicit incorporation of \(U(1)\) symmetry~\cite{liu2023supercomputing,liu2023simulating}, which naturally arises from conservation laws~\cite{singh2011tensor,singh2012tensor} such as particle number or total spin, into the MPS formalism. 

A quantum system is said to possess a global \(U(1)\) symmetry if its Hamiltonian \(\hat{H}\) commutes with a conserved charge operator \(\hat{N}\) as \([\hat{H}, \hat{N}] = 0\), such that the dynamics preserve the total charge. In the conventional canonicalized MPS representation~\cite{schollwock2011density}, 
the many-body wavefunction \(|\psi\rangle = \sum_{i_1,\ldots,i_M} c_{i_1,\ldots,i_M} |i_1,\ldots,i_M\rangle\) is approximated by factorizing the coefficient tensor \(c_{i_1,\ldots,i_M}\) as \(c_{i_1,\ldots,i_M} =
\sum_{\{\alpha\}} \Gamma^{[1] i_1}_{\alpha_0 \alpha_1} \lambda^{[1]}_{\alpha_1}
\cdots 
\Gamma^{[M] i_M}_{\alpha_{M-1}\alpha_M}\), where \(i_k\) denotes the local physical index at site \(k\) (dimension \(d\)), 
\(\alpha_k\) are the virtual bond indices with maximum dimension \(\chi\), 
\(\Gamma^{[k] i_k}_{\alpha_{k-1}\alpha_k}\) are site tensors carrying the physical degrees of freedom, 
and \(\lambda^{[k]}_{\alpha_k}\) are the Schmidt coefficients encoding bipartite entanglement between sites. 
The memory cost is dominated by storing the \(\Gamma\) tensors, which scales as \(\mathcal{O}(\chi^2)\). 

For systems that conserve a global charge, such as particle number or total spin, the wavefunction possesses a global \(U(1)\) symmetry. 
This symmetry can be incorporated directly into the MPS formalism, leading to a more compact representation and lower computational complexity. The \(U(1)\)-symmetric MPS~\cite{liu2023supercomputing,liu2023simulating} introduces charge indices \(c^{[k]}_{\alpha_k}\) that encode the number of particles to the right of each bond. 
The amplitude tensor becomes \( c_{i_1,\ldots,i_M} = 
\sum_{\{\alpha\}} \Gamma^{[1]}_{\alpha_0 \alpha_1}\lambda^{[1]}_{\alpha_1}\cdots
\Gamma^{[M]}_{\alpha_{M-1}\alpha_M}
\prod_{k=1}^M \delta(c^{[k-1]}_{\alpha_{k-1}} - c^{[k]}_{\alpha_k} - i_k) \), where \(c^{[k]}_{\alpha_k}\) is defined as the right-charge on bond $k$ , \ie, the total number of particles
to the right of site $k$ denoted as \(c^{[k]}_{\alpha_k} \;\equiv\; \sum_{j=k+1}^{M} i_j\). With this definition, the Kronecker deltas in the $U(1)$-symmetric MPS
enforce charge conservation across each site via \(c^{[k-1]}_{\alpha_{k-1}} - c^{[k]}_{\alpha_k} \;=\; i_k\), with boundary conditions $c^{[0]}_{\alpha_0}=N$ (total particle number)
and $c^{[M]}_{\alpha_M}=0$. As a result, the \(\Gamma\) tensors no longer carry explicit physical indices, the memory cost is reduced by a factor of the local dimension \(d\), and local updates require only block-sparse SVDs on submatrices \(\Theta(c^{[k]})\)~\cite{liu2023supercomputing,liu2023simulating}, each of size at most \(\chi \times \chi\) instead of a full \(\chi d \times \chi d\) matrix. 

Complementing this development is an efficient CPU and GPU implementation. On CPUs, the algorithm loops over allowed charge values to construct \(\Theta(c^{[k]})\) matrices. 
On GPUs, hierarchical tiling strategies at the thread, warp, and block level are combined with charge-sorted memory alignment to ensure coalesced access patterns and efficient use of tensor cores. At the distributed level, independent two-site updates and distinct \(\Theta(c^{[k]})\) blocks are processed in parallel across GPUs and nodes. 
This hierarchical parallelization enables strong scaling to modern supercomputers. \cite{liu2023supercomputing,liu2023simulating} benchmark on the Polaris system, showing nearly three orders of magnitude speedup compared to optimized CPU implementations.

\section{Beyond Symmetry-guided Principles and quantum simulation}

\textbf{Symmetry beyond \(U(1)\) in tensor networks for machine learning.} While the incorporation of $U(1)$ symmetry has proven powerful for quantum many-body simulations, 
the underlying idea of embedding symmetry into tensor network structures is not limited to $U(1)$ but extends naturally to more general symmetries such as $SU(2)$ or $O(3)$, and even beyond quantum physics into domains like machine learning: \cite{li2024unifying} proposed using \emph{fusion diagrams}, a technique widely employed in simulating $SU(2)$-symmetric quantum many-body systems, to design new spatially equivariant components for neural networks, referred to as \emph{fusion blocks} which act as universal approximators for equivariant functions. They ensure that the resulting architectures respect $O(3)$ symmetry while retaining expressive power. 
When incorporated into established models such as Cormorant~\cite{anderson2019cormorant} and MACE~\cite{batatia2022mace}, this approach achieved state-of-the-art accuracy in molecular property prediction benchmarks including QM9~\cite{ramakrishnan2014quantum} and MD17~\cite{chmiela2017machine}, as well as in molecular dynamics tasks such as the photoisomerization of stilbene. This highlights how efficient tensor network schemes in quantum many-body physics can inform the principled design of efficient, symmetry-preserving neural architectures.

\textbf{Symmetry-preserving variational quantum algorithms.} Beyond classical tensor network simulations, symmetry has also been exploited in the design of variational quantum algorithms~\cite{cerezo2021variational}, where incorporating group structure directly into circuit ansätze can improve efficiency.  \cite{zheng2023speeding} developed a theoretical framework for 
\(S_n\)-equivariant convolutional quantum circuits with global \(SU(d)\) symmetry, 
significantly generalizing Jordan's Permutational Quantum Computing (PQC)~\cite{jordan2009permutational} into a broader PQC+ formalism~\cite{zheng2022super}. 
By exploiting Schur--Weyl duality~\cite{childs2007weak,harrow2005applications,keppeler2018birdtracks} between \(SU(d)\) and \(S_n\) representations on \(n\)-qudit Hilbert spaces 
and employing the Okounkov--Vershik approach~\cite{okounkov1996new} with Young--Jucys--Murphy elements, 
they introduced the so-called \(S_n\)-equivariant Convolutional Quantum Alternating Ans\"atze (Sn-CQA). 
These ans\"atze provably generate any unitary within a given \(S_n\) irrep sector, establishing a restricted universality under global \(SU(d)\) symmetry. Moreover, the results provide a new proof of the universality of Quantum Approximate Optimization Algorithm~(QAOA)~\cite{farhi2014quantum,lieb1962ordering,cerezo2021variational} and show that 4-local \(SU(d)\)-symmetric unitaries are sufficient while 2-local ones fail for \(d \geq 3\) in building generic \(SU(d)\) symmetric quantum circuits. 
Numerical simulations demonstrated the ability of Sn-CQA 
to efficiently approximate ground states in regimes where classical tensor network and neural network quantum state ans\"atze struggle.

In sum, Sn-CQA exemplifies how embedding symmetry at the circuit-design level can both guarantee expressive completeness within symmetry sectors and restrict the variational search space, paralleling the role of symmetry-enhanced tensor network method in classical simulation. 
Besides, these suggest a broader direction: whether in tensor network contractions on supercomputers or in variational quantum algorithms, symmetry acts as a unifying principle enabling scalable quantum simulation. This further indicates that respecting the symmetry in variational ansätze is advantageous, as it might enables more efficient quantum simulation~\cite{vieijra2021many,vieijra2020restricted}.

\textbf{Beyond symmetry-guided principles in tensor networks.} While symmetry provides a unifying framework for scalable quantum simulation, machine learning, and variational quantum algorithms, there also exist approaches that pursue efficiency from different guiding principles. For instance, \cite{yuan2021quantum} proposes a framework of hybrid tensor networks. They combine classically contractable tensors and prepared quantum states to efficiently represent the quantum wave functions and simulate systems larger than conventional quantum simulation
algorithms~\cite{o2017quantum} in current quantum hardwares. 

Besides, beyond both symmetry-guided principles and quantum simulations, \cite{wei2025state} highlights how similar principles can be extended to broader ranges of quantum information processing. \cite{wei2025state} introduces circuits with new family of layouts termed as parallel-sequential (PS) circuits that interpolate between brickwall and sequential circuits, enabling efficient preparation of MPS with tunable entanglement and correlation range. PS circuits suppress error proliferation and exhibit superior noise robustness through balancing depth and gate count on NISQ quantum devices, illustrating how symmetry-inspired structures can guide the design of shallow, trainable variational ans\"atze for \textit{quantum state preparation}  task. 

\section{Conclusion}

In sum, symmetry-aware tensor networks deployed on modern supercomputers provide a practical path towards simulating larger and more complex quantum systems. At the same time, symmetry-inspired structures in machine learning and variational quantum algorithms promise new architectures and ans\"atze informed by the same design principles. Beyond symmetry, complementary approaches such as hybrid tensor networks and parallel–sequential circuits illustrate that efficiency can also arise from alternative guiding principles, for example by balancing classical and quantum resources or by optimizing circuit layouts for noise robustness. Together, these developments point toward a broader landscape of strategies rooted in both symmetry and symmetry-beyond that advance the long-term goal of scalable and efficient quantum simulation, quantum computation and machine learning.

\section*{Acknowledgments}
MC and JL are supported in part by the University of Pittsburgh, School of Computing and Information, Department of Computer Science, Pitt Cyber, Pitt Momentum fund, PQI Community Collaboration Awards, John C. Mascaro Faculty Scholar in Sustainability, NASA under award number 80NSSC25M7057, and Fluor Marine Propulsion LLC (U.S. Naval Nuclear Laboratory) under award number 140449-R08. This research used resources of the Oak Ridge Leadership Computing Facility, which is a DOE Office of Science User Facility supported under Contract DE-AC05-00OR22725. JL is also supported in part by International Business Machines (IBM) Quantum through the Chicago
Quantum Exchange, and the Pritzker School of Molecular Engineering at the University of Chicago through the Air Force Office of Scientific Research (AFOSR) Multidisciplinary Research Program of the University Research Initiative (MURI) (Grant No. FA9550-21-1-0209).

\bibliographystyle{IEEEtran}
\bibliography{refs} 
\end{document}